\documentstyle[floats,epsf,epsfig,prbbib,aps]{revtex}

\begin{document}

\twocolumn[\hsize\textwidth\columnwidth\hsize\csname
@twocolumnfalse\endcsname

\title{Equilibrium shapes and energies of coherent strained InP islands}
 
\author{Q. K. K. Liu} 
\address{Bereich Theoretische Physik, Hahn-Meitner-Institut, Glienicker
  Str. 100 D-14109 Berlin, Germany}
\author{N. Moll and M. Scheffler} 
\address{Fritz-Haber-Institut der Max-Planck-Gesellschaft, Faradayweg 4-6,
  D-14195 Berlin-Dahlem, Germany} 
\author{E. Pehlke}
\address{Physik-Department T30, Technische Universit\"{a}t M\"{u}nchen,   
  D-85747 Garching, Germany}
\maketitle

\begin{abstract}
The equilibrium shapes and energies of coherent strained InP islands 
grown on GaP have been investigated with a hybrid approach that has 
been previously applied to InAs islands on GaAs. This combines 
calculations of the surface energies by density functional theory and 
the bulk deformation energies by continuum elasticity theory. The 
calculated equilibrium shapes for different chemical environments 
exhibit the \{101\}, \{111\}, \{\=1\=1\=1\} facets and a (001) top 
surface. They compare quite well with recent atomic-force 
microscopy data. Thus in the InP/GaInP-system
a considerable equilibration of the individual islands with respect to
their shapes can be achieved.
We discuss the implications of our results for the 
Ostwald ripening of the coherent InP islands. In addition we 
compare strain fields in uncapped and capped islands.
\end{abstract}

\vskip2pc]

\thispagestyle{empty}

\section{INTRODUCTION}

It was observed about ten years ago that during the
hetero-epitaxial growth of lattice-mismatched semiconductors small 
dislocation-free islands can form.\cite{ea90,mo90,gu90,sn91,or92}
To emphasize the zero-dimensional character of their electronic density 
of states, these objects have been labeled quantum dots. 
Since then, the quantum dots have attracted an enormous interest in the
area of semiconductor physics.\cite{bi:98} 
Besides being fascinating objects for basic research, there are also various 
potential applications, ranging from 
improved device properties of quantum-dot semiconductor-lasers, which have
in fact already been demonstrated,\cite{ki:94} up to more exotic
applications as part of a quantum computer.\cite{za:98}
The ``self-organized'' hetero-epitaxial growth of arrays of
quantum dots with a preferentially narrow size distribution 
is thus the aim of intense research. 
To finally be able to optimize the growth parameters, a detailed knowledge 
of the growth mechanism, its energetics and its kinetics, is essential.
 
There is general agreement about the main driving force for, e.g., 
InAs/GaAs or InP/GaInP  island formation in the Stranski-Krastanov 
growth-mode. The growth begins with the deposition of a highly strained 
two-dimensional film, the wetting layer. With the addition of more 
material beyond a critical thickness the film becomes meta-stable. 
Coherent, i.e., dislocation-free islands form due to the energy gained 
by strain relaxation in the islands.  However, the details of the growth 
mechanism\cite{he:97} are not yet well understood. In fact, there are 
competing theories, which are based on energy ground-state considerations 
on the one hand, or on kinetic, i.e.\ non-equilibrium effects on the 
other.\cite{pr95,ch96,je96,do97,te98} Accordingly, the final destiny of 
the islands, as predicted by the various theories, is rather different. 
The more conventional fate of the islands would be to undergo Ostwald 
ripening,\cite{os00,zi92} i.e., the larger islands would grow at the 
expense of the smaller ones. The resulting island size distribution would 
be rather broad. In fact, for the systems of concern in our study, 
Ostwald ripening takes place at a slower rate than the growth rate of 
the islands. Kinetic effects could be active that effectively decrease 
the growth rate of the larger islands, thus sharpening the size 
distribution.\cite{ch96,je96,se96} In contrast to these growth scenarios, 
Shchukin {\it et al.}\cite{sh95,da:97} have suggested that there exists 
a range of parameters for which islands of a finite size are 
thermodynamically stable. Our {\it ab initio} results for the surface 
energies and surface stresses, however, indicate that neither in the case 
of InAs/GaAs nor InP/GaInP would this mechanism result in the formation 
of stable islands.

For the most widely studied system of InAs islands on GaAs, a range of 
growth parameters seem to have been established for producing islands of 
certain uniform densities and a rather narrow size 
distribution,\cite{mo94,le94,ru95,bi:98}\ although these growth parameters 
are still subjects of discussion in the literature. Spectroscopic studies 
of the islands, after they were capped by barrier materials, have been 
reported.\cite{gr95a,sc96,he96a,sa97,mi97}\ To investigate the 
equilibrium shape and stability of the coherent strained InAs islands 
at low island densities, we have applied a hybrid method to calculate 
the total energy.\cite{mo96,pe97,mo98}\ In this approach the energy 
gained by island formation is decomposed in the following form:
\begin{equation} 
E^{\rm total}  =  E^{\rm relax} + E^{\rm surface} + E^{\rm edge} ,
\end{equation} 
where $E^{\rm relax}$ is the gain in deformation energy when the 
material forms a strained island instead of a biaxially strained film, 
$E^{\rm surface}$ is the cost in surface energy due to the creation of 
facets on the sides of the island instead of the surface covered by 
the base of the island, and $E^{\rm edge}$ is the energy cost for  
the creation of sharp edges. For an isolated island to form at all in 
preference to a film, $E^{\rm total}$ must be negative. The surface 
energy $E^{\rm surface}$ is calculated {\it ab initio}, applying 
density-functional theory (DFT). In Ref.\onlinecite{pe97} the surface 
energies $E^{\rm surface}$ corresponded to those of unstrained 
surfaces. This approximation was subsequently improved upon in 
Ref.\,\onlinecite{mo98}, in which the renormalization of the surface 
energies due to surface stress was taken into account. However, the 
corrections amounted to a reduction of the surface energies by at most 
11\,\% and left the prediction for the equilibrium shape qualitatively 
unchanged. Furthermore, the edge energy $E^{\rm edge}$ was estimated 
to be negligible, provided the island size is not too small. The 
elastic strain field can be treated within a continuum approach. Thus 
$E^{\rm relax}$ is calculated within linear elasticity theory using a 
finite-element method (FEM). The effect of nonlinearity was seen to be 
small. 

As a result, we obtained a volume-dependent optimum shape for the InAs 
islands, which can be described as a (001)-truncated pyramid with 
\{101\}, \{111\}, and \{1\=11\} faces. However, the diversity of 
experimentally observed island shapes appears not to be reconcilable 
with the assumption of thermodynamic equilibrium. Among the 
experimental results, the mostly square-based islands have 
faces \{101\},\cite{ru95}\ \{105\},\cite{so96a}\ \{113\},\cite{mo94}\ 
\{136\},\cite{le98}\ and a series of islands of low aspect ratios whose 
morphologies change according to the coverage.\cite{le94}\ We take this 
difference as an indication that a sensitivity to growth conditions 
and kinetic effects have to be featured in a growth theory, including 
the possibility that the deposited material may migrate or segregate as 
witnessed in the growth of quantum wells,\cite{mu92,le95} and 
self-organized islands.\cite{xi94,gr97,ga97}

While the predictions of the hybrid method for the equilibrium shape of 
InAs/GaAs coherent islands have not been borne out so far in experiment, 
i.e., in the `window' of growth condition assumed, the experimental 
characterization of the three-dimensional islands in the InP/GaInP 
system,\cite{se96} briefly summarized in the next paragraph, seems to 
indicate a better chance for an equilibrium approach to be 
valid. Therefore, InP/GaInP represents an excellent benchmark system to 
show both the applicability of the hybrid method and the notion that 
the shape equilibration of coherent islands can be achieved.

The growth of InP/GaInP by metal-organic vapor phase epitaxy (MOVPE) has
consistently yielded uncovered 
\mbox{islands \cite{se96} that are significantly} 
larger \mbox{($\sim 45 \times 60$ nm$^2$)} than their counterparts in 
InAs/GaAs \mbox{($\sim 12 \times 12$ nm$^2$)} grown by molecular beam 
epitaxy (MBE). Transmission electron microscopy (TEM) and atomic force 
microscopy (AFM) have yielded strong evidence that they have facets 
of low Miller indices only,\cite{ge95,pi96} rather reminiscent of the 
previous prediction for InAs/GaAs. They are stable against annealing 
\cite{se96} of several minutes at the growth temperature of 
580 $^{\circ}$C, and their morphology has been reported to remain 
unchanged after overgrowth with capping material.\cite{ca98}\ Although 
for these materials there is a lack of specific observations of 
segregation inside the islands and diffusion between the islands and the 
barrier, it is probable that both takes place. To what extent they affect 
the shape and size of InP/GaP islands remains to be determined. It is 
also becoming clear that the spectroscopy of the islands depends not only 
on the volume but also significantly on the strain distribution inside 
the islands, and the latter is greatly influenced by the shape of the 
islands.\cite{ki98a,pr97}\ The results for InP/GaInP complement those 
for InAs/GaAs, e.g., growth of quantum wells on InP islands, \cite{ah94} 
growth characteristics,\cite{ge95,ku95,re95,re96,ju98}
photoluminescence,\cite{ah94,ku95,pi95,ca95,an95,he96b} optical gain 
and lasing,\cite{mo96a} InP islands used as stressors to induce 
quantum dots in a quantum well,\cite{so95,tu95,so96} Landau levels 
formation in InP islands,\cite{no97} theoretical study of the electronic 
states of the islands.\cite{pr97,fu97} 

We are thus encouraged to conduct a study of InP/GaInP-islands parallel 
to our previous work for InAs/GaAs, and hence enlarge our general 
understanding of coherent strained islands. Furthermore, gaining an 
insight into the strain distribution of capped islands would help 
towards a better understanding of the potential experienced by the 
electrons and holes at the quantum dot and in its vicinity. Of course, 
if the effect of phase segregation inside the island or diffusion of 
atoms to and from the barrier turns out to be substantial, our approach 
would need to be generalized. In any event, our findings could still be 
used as the starting point for a yet more realistic modeling.

The organization of the work is as follows: First, we present the DFT 
results for the surface energies. In Section III, we describe briefly 
our FEM simulations and derive the equilibrium shape. We shall indicate 
how Ostwald ripening emerges from our model, assuming that no other 
faster kinetic processes have preempted its time development. In 
Section IV we compare the strain distributions of uncapped and capped 
islands. The implication for the use of strained islands as a stressor 
for quantum wells will be pointed out. A generalization from InP/GaP to 
InP/Ga$_{\rm x}$In$_{\rm 1-x}$P islands, i.e., to systems with a 
different lattice mismatch, is included in the final Section V.

\section{SURFACE ENERGY}

As in the case of InAs we assume that the relevant surface 
reconstructions, i.e., those of lowest energy, correspond to the low-index 
surface orientations \{110\},\{100\},\{111\}, and \{\=1\=1\=1\}. 
The details of the calculation of surface energies by DFT are described 
in Refs.\,\onlinecite{mo96,mo98}. 
We apply the local density approximation to the 
exchange-correlation energy-functional and use {\it ab initio} 
norm-conserving, fully separable pseudopotentials. The plane-wave 
expansion of the wave function has an energy-cutoff of 10 Ry, and the 
{\bf k}-summation makes use of a uniform Monkhorst-Pack mesh with a density 
equivalent to 64 {\bf k}-points in the complete $(1 \times 1)$ surface 
Brillouin zone of the (100) surface. We neglect 
the correction of the surface energies due to strain, as, 
for the InAs/GaAs-islands, we found it to be small and not affecting our 
conclusions concerning the island shape and stability.\cite{mo98} 
 
\begin{figure}[tb]
  \begin{center}
    \epsfig{figure=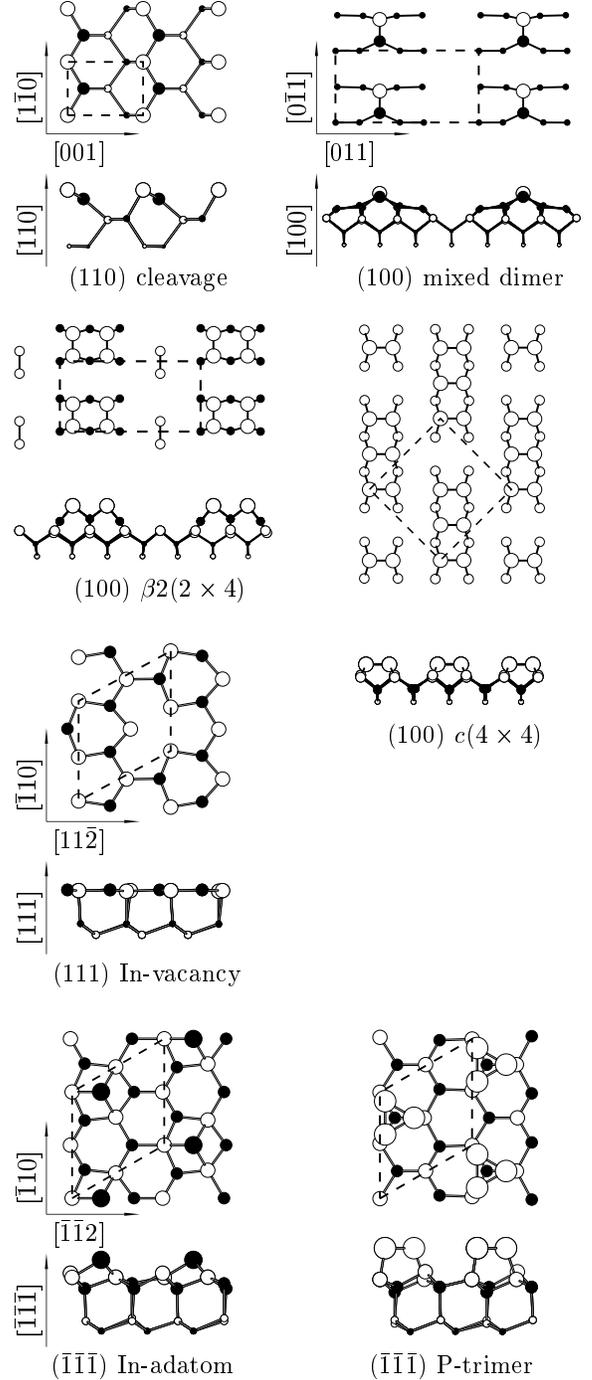, width=8cm}
  \end{center}
  \caption{Atomic structure models for the different InP surfaces, top
    and side views. Filled and open circles denote In and P atoms,
    respectively.}
  \label{fig:geometries}
\end{figure}

\begin{figure}[tb]
  \begin{center}
    \epsfig{figure=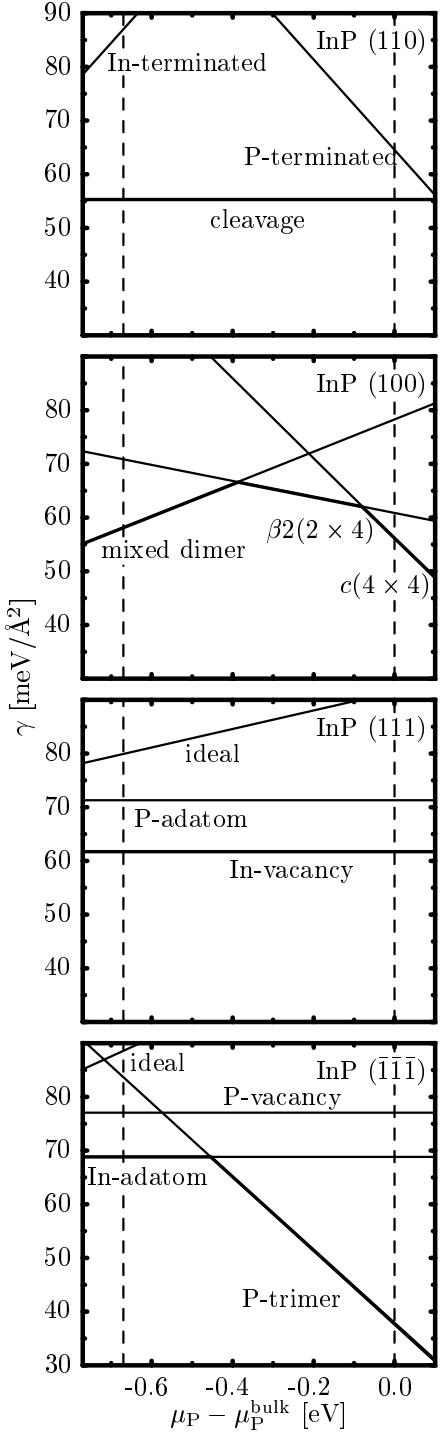, width=6cm}
  \end{center}
  \caption{InP surface energies of the (110), (100), (111), and (\=1\=1\=1)
    surface orientations as a function of the phosphorous chemical 
    potential. The thick
    lines highlight the calculated surface energies of the reconstructions 
    of lowest total energy.}
  \label{fig:energies}
\end{figure}

The surface atomic structures for different surface orientations 
are shown in Fig.\,\ref{fig:geometries}. 
The corresponding surface energies as a function of the phosphorous 
chemical potential are shown in Fig.\,\ref{fig:energies}, where the left 
and right vertical dashed lines mark the limits for 
In- and P-rich environments, 
respectively. They are characterized by the coexistence of the InP 
surface with either an In or a P bulk phase.
Since epitaxial growth takes place mostly in a P-rich environment, we 
list in Table I the surface energies of the stable reconstructions 
for the chemical potential 
$\mu_{\rm P} = \mu_{\rm P}^{\rm bulk} - 0.1$\,eV. It will be shown in 
Section V that the experimentally observed coherent islands are best 
compared with the theoretical results at this chemical potential.

There are a great deal of similarities between InP and InAs surfaces in 
equilibrium\cite{moe93,pe97,mo98} and some interesting differences. 
Both the (110) and (111) surface energies are independent of the phosphorous 
chemical potential. The $(1 \times 1)$ relaxed cleavage surface and 
the stoichiometric In vacancy structure are the stable reconstructions 
for the (110) and (111) orientations, respectively. The surface 
energies are 55 and 62 meV/${\rm \AA}^{2}$, respectively. These are to 
be compared with those of InAs of 41 and 42 meV/${\rm \AA}^2$. The 
cleavage surfaces of the (110) orientation in InP have been observed 
experimentally using low-energy electron-diffraction \cite{fo92} (LEED) 
and for both InP and InAs using low-energy positron-diffraction.\cite{ch93} 
A DFT study of InP(110) surface has also been carried 
out.\cite{um94} We find the equilibrium structure of the (\=1\=1\=1) 
surfaces to be the same as for InAs. On the (\=1\=1\=1) surface in 
P-rich environment, the P-trimer reconstruction is preferred, i.e., 
within some interval near the right-hand dashed line in 
Fig.\,\ref{fig:energies}.
\begin{table}[tb]
\caption{The relaxed surface reconstructions of InP for $\mu_{\rm P} = 
         \mu_{\rm P}^{\rm bulk} - 0.1$\,eV, and surface energies.}
\begin{tabular}{ccccc}
orientation & reconstruction            & surface energy \\ 
            &                           & [meV/${\rm \AA}^{2}$] \\ \tableline
(110)       & cleavage                  & 55             \\
(100)       & $\beta 2(2 \times 4)$     & 62      \\ 
(111)       & $(2 \times 2)$ In-vacancy & 62  \\ 
(\=1\=1\=1) & $(2 \times 2)$ P-trimer   & 44  \\
\end{tabular}
\end{table}
At $\mu_{\rm P} = \mu_{\rm P}^{\rm bulk} - 0.1$ eV the InP (\=1\=1\=1) 
surface energy is 44 meV/${\rm \AA}^{2}$ compared to 36 
meV/${\rm \AA}^{2}$ for InAs. A $(\sqrt{19} \times \sqrt{19})$ 
reconstruction of the (\=1\=1\=1) surface \cite{bi90} seems not relevant 
for our present study, as the area on side faces of the island might be 
too small to accommodate this reconstruction without having to invoke 
edge effects. 

The (100) reconstructed surfaces of InP and InAs had been generally 
assumed to be similar, too. However, recent DFT studies and 
experimental observations by several groups using low-energy electron 
diffraction (LEED), reflection anisotropy spectroscopy (RAS), soft 
x-ray photoelectron spectroscopy (SXPS), and scanning tunneling 
microscopy (STM) have yielded evidence that there is a qualitative 
difference between the atomic structure of the InP and the InAs(100) 
surfaces.\cite{ki98b,pa98,sc98}\ In a very P-rich environment, i.e., 
at $\mu_{\rm P}$ rather close to $\mu_{\rm P}^{\rm bulk}$, DFT predicts 
the $c(4 \times 4)$ reconstruction to be more stable than the 
$\beta 2(2 \times 4)$ reconstruction, which is not the case for InAs. 
For InP in moderately P-rich and InAs in As-rich environments, 
$\beta 2(2 \times 4)$ is the common stable reconstruction. However, in 
an In-rich environment InP displays a mixed dimer reconstruction, in 
contrast to the $\alpha (2 \times 4)$ reconstruction for InAs. However, 
for the growth of strained islands in a moderately P-rich atmosphere, 
it is still the $\beta 2(2 \times 4)$ reconstruction that is selected 
as in InAs/GaAs, with the surface energy equal to 
62 meV/${\rm \AA}^{2}$ compared to 44 meV/${\rm \AA}^{2}$ for InAs.

\section{EQUILIBRIUM ISLAND SHAPES}

For a fixed volume of the island (or, equivalently, a fixed number of 
atoms), the equilibrium shape minimizes the total energy of the system, 
$E^{\rm total}$ of Eq.\,(1), with respect to all possible shapes. As 
explained above, the surface energy $E^{\rm surface}$ is calculated for 
the unstrained surface and the edge energy $E^{\rm edge}$ is 
neglected.\cite{mo98}\ For a square-based island bounded solely by 
the four \{101\} surfaces, the length of the base is chosen equal to 
12.9 nm and the height to 6.5 nm. This determines the volume that remains 
unchanged for all calculated islands. The island dimensions vary only 
moderately for truncated islands bounded by the other low-index surface planes 
we have included in this study. By construction the islands are placed 
40 nm apart, and reside on a substrate of thickness 24 nm. 
As in the case of InAs,\cite{pe97,mo98} 
the thickness of the wetting layer has been set to zero.

We have used the commercial product MARC \cite{ma96} to perform the 
finite-element simulations. One notable feature of all commercial 
products is that the preferred finite element (FE) is an eight-node 
hexahedron or three-dimensional arbitrarily distorted cube. 
It has been shown that 
it is superior to the simple tetrahedron in terms of fast convergence 
and computing speed. We have also adopted this class of FE for all our 
calculations, with the understanding that corners of the hexahedron 
can merge to form wedge-shaped or pyramid-shaped elements. 
 
Throughout this work we have taken InP to be the island and wetting 
layer material, and GaP to be the barrier material, or the substrate 
when the island is uncapped. The corresponding experimental lattice 
constants and elastic moduli \cite{la79} are listed in Table II. 
\begin{table}[b]
\caption{The experimental lattice constants $a$ and elastic moduli 
$c_{11}, c_{12}$ and $c_{44}$ of InP and GaP.} 
\begin{tabular}{ccccc}
    & $a$ \AA & $c_{11}$ [GPa] & $c_{12}$ [GPa] & $c_{44}$ [GPa] \\ \tableline
InP & 5.87    & 102            & 58             & 46             \\
GaP & 5.45    & 141            & 63             & 72             \\
\end{tabular}
\end{table}
The lattice mismatch is 
$\alpha := (a_{\rm GaP}-a_{\rm InP})/a_{\rm InP} = -7.1\,\% $.
Although this mismatch is quite large, we restrict ourselves to linear 
elasticity theory. 
Taking the elastic moduli from Table II, the deformation energy per unit 
volume of the biaxially strained 
uniform InP film amounts to 3.0 meV/$\rm {\AA}^3$.

In our FEM calculations of $E^{\rm relax}$ the number of FE 
varies slightly, depending on the island's bounding surfaces. The number 
of FE is increased in the areas where the deformation energy density is 
large until a 1\,\% accuracy is achieved. In general, there are 
approximately 7000 FE in total, of which approximately 2000 are 
distributed in the island. For volume-conserving truncated islands, the 
elastic energy can be accurately approximated from the untruncated 
value by an analytical formula.\cite{pe97} 

The anisotropy of the surface energy driving the formation of particular 
crystal facets is an essential aspect of our approach.\cite{sp:97}
We have investigated island shapes which are bounded by the low 
Miller-index surfaces described in Section II. The collection of shapes 
we have considered is the same as in the study of the InAs 
islands.\cite{pe97,mo98} The island base has the orientation (001) 
and the top of the island may be truncated by a plane of the same 
orientation (see Fig.\,6 of Ref.\,\onlinecite{mo98}). 

\begin{figure}[tb]
  \begin{center}
    \epsfig{figure=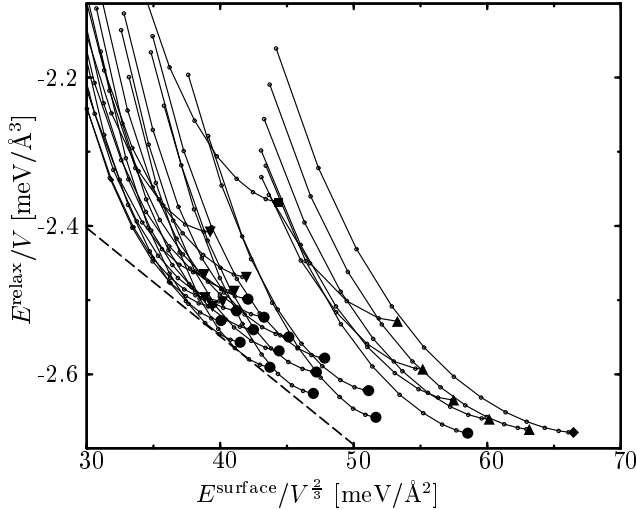, width=\linewidth}
  \end{center}
  \caption{The relaxation energy per unit volume $E^{\rm relax}/V$ versus
    $E^{\rm surface}/V^{2/3}$ for InP islands. 
    {\sl Square} : square-based island with four \{101\}
    facets. {\sl Diamond} : square-based island with two \{111\} and two
    \{\=1\=1\=1\} facets. {\sl Triangles up} :
    huts with two \{111\} and two \{\=1\=1\=1\}
    facets.  {\sl Triangles down} : square-based \{101\} island with
    \{\=1\=1\=1\} truncated edges.  {\sl Dots} :
    islands with four \{101\}, two \{111\}, and two
    \{\=1\=1\=1\} facets. The dashed line is the
    line of constant total energy $E^{\rm relax} + E^{\rm surface}$ = constant
    that selects the equilibrium shape for the volume 
    $V = 3 \times 10^{5} {\rm \AA}^{3}$. See text for further explanations.}
  \label{fig:optimization}
\end{figure}

The results of energy evaluated according Eq.\,(1) are summarized in a 
scale-invariant manner in Fig.\,\ref{fig:optimization}, 
where the ordinate $x = E^{\rm surface}/V^{2/3}$ and abscissa 
$y = E^{\rm relax}/V$. 
We have taken the surface energies at the phosphorous chemical potential 
$\mu_{\rm P}$ = $\mu_{\rm P}^{\rm bulk} - 0.1$ eV. The solid symbols 
denote untruncated islands for which a full FE calculation has been 
carried out. Qualitatively these results are similar to those of 
InAs/GaAs.\cite{pe97,mo98}\ We find that a square-based island with 
\{101\}-facets only (solid square) has a larger bulk deformation energy 
than a square-based island with two \{111\}- and two \{\=1\=1\=1\}-facets 
(rhombus). The latter has steeper side facets which allow a more efficient 
stress relaxation. The line that emanates from each of the solid symbols 
joins up 
the small dots for which $E^{\rm relax}$ was derived from the analytical 
approximation for the volume-conserving truncated island. For a given 
volume $V$, islands with the same total energy $E^{\rm total}$ lie on the 
straight line 
\begin{equation}
 \frac{E^{\rm total}}{V} =
                    \frac{E^{\rm relax}}{V} + \frac{E^{\rm surface}}{V}
                         = y + V^{-1/3} \cdot x = {\rm constant} ,
\label{etotbyv}
\end{equation}
plotted as a dashed line in Fig.\,\ref{fig:optimization}. 
(Note that the edge energies $E^{\rm edge}$ have been neglected.)
It is clear from Eq.\,(\ref{etotbyv}) that 
the volume of the island $V$ is related to the negative slope of the 
line.
For a given volume $V$, the equilibrium shape of the island is 
determined by the first point of contact from below of the straight line, 
Eq.\,(\ref{etotbyv}), with the calculated island-energy curves. 
 
For a given island shape, $E^{\rm relax}$ and $E^{\rm surface}$ scale
like $V$ and $V^{2/3}$, respectively.
This means that the data in Fig.\,\ref{fig:optimization} are invariant 
against a change of the volume of all islands by the same
factor.  However, the slope of the straight line, Eq.\,(\ref{etotbyv}), 
changes, and hence a 
different equilibrium shape will be selected. By inspection, the
results of Fig.\,\ref{fig:optimization} indicate that for larger V, i.e., 
smaller negative slope, untruncated islands are most likely to be 
selected, while smaller $V$ favors truncated islands as equilibrium 
shape. The interplay of $E^{\rm relax}$ 
and $E^{\rm surface}$ in Eq.\,(1), and how their relative weightings 
change as a function of the island volume have been discussed and 
supported by explicit calculations in Ref.\,\onlinecite{pe97}. 
Fig.\,\ref{fig:optimization} serves as a compact way of encapsulating 
the result for the equilibrium shapes for arbitrary volumes, provided 
that the volume (or the number of atoms) of the island is not too small 
to invalidate our hybrid ansatz, Eq.\,(1), or 
too large such that the island is no longer coherent (dislocations 
appear).

\begin{figure}[tb]
  \begin{center}
    \epsfig{figure=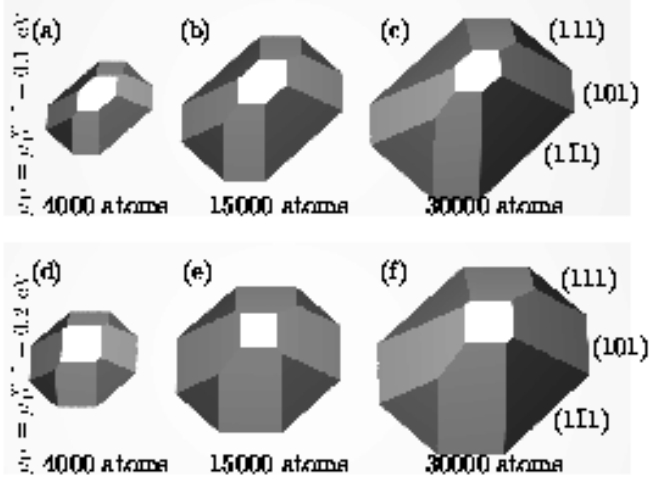, width=\linewidth}
  \end{center}
  \caption{The equilibrium shapes of coherent strained InP islands for 
    three different volumes and two different chemical potentials. The 
    unstrained volumes in ascending order are approximately 
    $9 \times 10^{4} {\rm \AA}^{3}$, $3 \times 10^{5} {\rm \AA}^{3}$, 
    $6 \times 10^{5} {\rm \AA}^{3}$.}
  \label{fig:shapes.eps}
\end{figure}

We display in Fig.\,\ref{fig:shapes.eps} examples of 
equilibrium shapes for three different island volumes.
They illustrate nicely that for a smaller volume a relatively larger amount of
material is truncated from the top of the island.  Additionally, we can
also inspect the difference in shapes due to a variation of the chemical potential:
The \{1\=11\}-facet, which is favored in a P-rich environment, is 
consistently more prominent for 
$\mu_{\rm P}$ = $\mu_{\rm P}^{\rm bulk} - 0.1$\,eV than for 
$\mu_{\rm P}$ = $\mu_{\rm P}^{\rm bulk} - 0.2$\,eV. 
The equilibrium shapes shown in 
Fig.\,\ref{fig:shapes.eps} have much similarities with the experimentally 
observed islands,\cite{ge95,pi96} in particular all observed facets are 
well accounted for by our theory.
Note that the experiments have been carried out for InP/GaInP instead of
InP/GaP, thus the experimental and theoretical lattice mismatch differ by 
nearly a factor of two. 
However, as will be discussed below, this can be accounted for by a simple
re-scaling of the island volumes.

In the theory of Shchukin {\sl et al},\cite{sh95} a combination of 
material parameters determines whether thermodynamic equilibration 
drives the system towards a stable array of islands or whether Ostwald 
ripening occurs. A stable array would arise,
if $E^{\rm surface}$ in Eq.\,(1) is modified by the surface 
strain in such a way that it changes sign. Making use of our 
quantitative surface energies, we can estimate, whether this condition 
is likely to be fulfilled in case of InP or not.

The change in surface energy $E^{\rm surface}$ is given by 
a sum over all facets of the island:
\begin{eqnarray}
 E^{\rm surface} & = & \sum_{i}^{\rm facets} \gamma^{i} \cdot A^{i}
                     - \gamma^{\rm base} \cdot A^{\rm base} \nonumber \\
                 & = & A^{\rm base} 
                       \sum_{i}^{\rm facets} 
                       \left[ \frac{\gamma^{i}}{\cos \theta^{i}} 
                     - \gamma^{\rm base} \right] \cdot \nu^{i}
\label{sum}
\end{eqnarray}
$A^{\rm base}$ and $A^{i}$ denote the areas of the base and the $i$-th 
facet on the surface of the island, respectively, $\theta^{i}$ is the 
angle between the $i$-th facet and the substrate surface, and $\nu^{i}$ 
is the ratio of $A^{i}$ projected onto the base and $A^{\rm base}$. The 
$\gamma^i$ are the surface energies from Table I. Each term within the 
summation of Eq.\,(\ref{sum}) is positive, and hence $E^{\rm surface}$ 
is positive. This has the consequence that, for a fixed equilibrium 
island shape, the volume derivative of Eq.\,(\ref{etotbyv}) is always 
negative, i.e., the total energy per unit volume decreases as the 
volume increases, thus favoring larger and larger islands, leading to 
Ostwald ripening. To change the sign of the respective individual 
contribution to Eq.\,(\ref{sum}), a strain-induced renormalization of 
the surface energy of approximately 20\,\% for the facets with (101) 
and (\=1\=1\=1) orientations, and 40\,\% for the facet with (111) 
orientation would be necessary. For the case of InAs/GaAs, the 
renormalization amounted to at most 11\,\%.\ \cite{mo98}\ Therefore, 
it seems reasonable to argue that Ostwald ripening is also the fate 
of the InP islands. Furthermore, a majority of the InP islands are grown 
on GaInP-substrates  whose lattice mismatch is about 50\,\% smaller than 
in the InP/GaP system. This further diminishes the importance of the 
renormalization of the surface energy due to the strain field.
Altogether this discussion seems to indicate that we are outside the 
parameter range for which arrays of stable islands occur. It is worth 
noting that the parameter range that conforms to 
Shchukin {\sl et al},\cite{sh95} implies that $E^{\rm surface}$ in 
Eq.\,(1) changes sign due to stress, i.e., one has the unusual situation 
that there is a gain in total energy by creating new surfaces. 
Furthermore, Shchukins model includes edge energies and island-island 
interaction to prevent the islands from coalescing. In 
comparison, a new method\cite{wa99} to tackle this problem of the 
stability of the islands has a much less restrictive ansatz and the 
agreement with experiments is encouraging.

\section{COMPARISON OF UNCAPPED AND CAPPED ISLANDS}

In this section we compare the strain fields for capped and 
uncapped islands. It is well documented that during 
overgrowth of GaAs over InAs islands there is significant migration of 
atoms as well as change in the morphology of the 
islands,\cite{xi94,ga97} and that In segregates during the overgrowth of 
GaAs over InGaAs islands or quantum wells.\
\cite{mu92,le95,gr97,na93,gr96,wo97}\ A well-studied sample of InP 
strained islands was grown on InGaP lattice-matched  
to a GaAs substrate.\ \cite{ge95,ca98}\ These InP islands seem quite 
stable - when uncapped, they survive a few minutes of annealing at 
580 $^\circ$C without changes and retain their shapes after overgrowth. 
These may be signs that the MOVPE-grown InP islands are less affected 
by atom migration and In segregation during overgrowth than the 
MBE-grown InAs islands. Therefore, until more specific observations of 
these effects are available, it would be interesting to compare the 
strain distributions of uncapped and capped islands without taking into 
account any change of shape or In segregation. While we are aware that 
the present results may be subjected to revision, we shall discuss 
below some rather general effect of the capping layer on the strain 
distributions.

We have adopted the shape and size of the island that had been 
observed by Georgsson {\sl et al},\cite{ge95} displayed in their 
Fig.\,1, i.e., a truncated pyramid grown on the (001) 
plane, with the defining planes for the side walls being (011), (101),
(1\=11), (0\=11), (\=101), and (\=111). The base is elongated in the [110] 
direction and has a length of 66 nm, a width of 40 nm in the 
[\=110] direction, and the height is 18.4 nm. 
The island resides on a 0.6 nm thick wetting layer. 
By construction the nearest-neighbor islands are 140 nm apart along 
the [110] and [\=110] directions. 
The size of our cell is chosen large enough such that away from 
the island and the wetting layer the barrier material is not 
significantly strained.
Since the island has reflection symmetry across two planes, only one 
quarter of the unit cell needs to be explicitly considered in the 
simulation. In the case of the uncapped island we have used in total 
1872 FE, while for the capped island we used 3756 FE. 
Our simulation of InP/GaP is 
not exactly a reproduction of the InP/GaInP coherent island in 
Ref.\,\onlinecite{ge95}, 
because the lattice mismatch of the latter was smaller, 
$\alpha \approx 3.7\,\%$. We discuss the scaling 
with respect to the lattice mismatch at the end of this section.

\begin{figure}[tb]
  \begin{center}
    \epsfig{figure=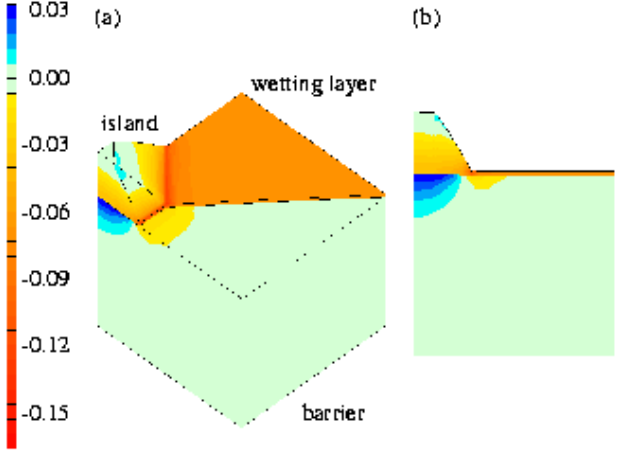, width=\linewidth}
  \end{center}
  \caption{Strain distribution of an uncapped truncated InP island 
    on a wetting layer and a GaP substrate.}
  \label{fig:uncappedf}
\end{figure}

We display in Fig.\,\ref{fig:uncappedf} two views 
of the trace of the strain tensor field, 
$\epsilon_{11}+\epsilon_{22}+\epsilon_{33}$, 
for an uncapped island. 
In Fig.\,\ref{fig:uncappedf} (a) a part of the wetting layer has been 
peeled off to reveal the strain distribution in the substrate.
Qualitatively the results are quite similar to those for InAs/GaAs 
without the wetting layer.\ \cite{mo98}\ One notes that the unhindered 
relaxation of the island in the upward direction has produced a small 
area of tensile strain at the top of the island. This seems to be a 
common result in all our calculations, whether the island is truncated 
or not. In the substrate, the strain distribution is characterized by 
compressive strain around the edges of the island, and tensile strain 
directly underneath the island. Such contrasting strains created by 
coherent strained islands of InP have been utilized as stressor to 
achieve three-dimensional quantum confinement \cite{so95,tu95,so96} 
in sub-surface
quantum-well structures, and a theoretical strain analysis similar to 
ours by FEM has been reported. In this discussion, an InGaAs 
quantum well (QW) is sandwiched between GaAs, with self-organized InP 
islands grown on top. The strain distribution thus produced modulates 
the conduction band of the QW, and creates a potential well. 
Photoluminescence spectra of this type of potential well have been 
measured and analyzed, hence this object has also earned the label of 
quantum dot. 
   
\begin{figure}[tb]
  \begin{center}
    \epsfig{figure=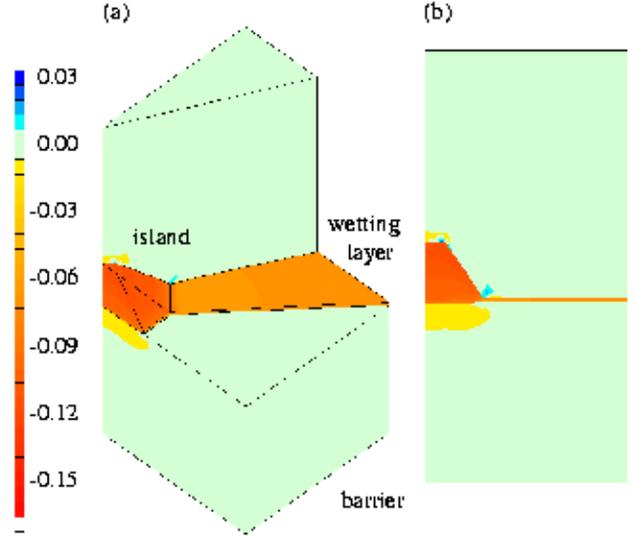, width=\linewidth}
  \end{center}
  \caption{Strain distribution of a truncated InP island on a wetting 
    layer, embedded in GaP.}
  \label{fig:cappedf}
\end{figure}

For comparison we display in Fig.\,\ref{fig:cappedf} two views of a 
capped InP island. 
In Fig.\,\ref{fig:cappedf} (a) part of the capping material is removed 
to reveal the island and the wetting layer. Part of the latter is 
also peeled off to show the substrate underneath. The island in this 
instance is entirely under compressive strain.\ \cite{gr95b}\ The area 
of tensile strain previously found in the substrate directly underneath 
the uncapped island has been replaced by an area of compressive strain. 
Small pockets of tensile strain are to be found in the vicinities of 
sharp corners in the capping layer. The disappearance of the pocket of 
tensile strain in the substrate has great significance for the creation 
of quantum dots in QWs as described above.  The amount of capping 
material affects the strain-induced potential well in the QW. This 
phenomenon is quite general and not limited to strained InP islands. 
If we assume that the strain interaction is given by a `hydrostatic' 
potential, i.e., $V \propto {\rm tr} \epsilon$, 
our present calculation shows the extreme case of destroying this 
potential well completely. 

A closer examination of the components $\epsilon_{11}$, $\epsilon_{22}$, 
and $\epsilon_{33}$ directly underneath the island reveals that 
$\epsilon_{11}$ and $\epsilon_{22}$, are tensile and remain 
qualitatively unchanged whether the island is capped or not. On the 
other hand, $\epsilon_{33}$ is slightly tensile for the uncapped island 
and overwhelmingly compressive for the capped island. These mechanical 
responses seem understandable. For capped or uncapped islands the cause 
of the biaxial tensile strain in the substrate perpendicular to the 
[001] direction is the same. However, the uncapped island heaves up in 
the [001] direction because the relaxation in this direction is 
unhindered. For the capped island, the capping material prevents the 
unrestrained heaving in the [001] direction. Hence the island exerts a 
considerable pressure onto the substrate as well, i.e., material in the 
substrate is compressed. Therefore the thickness of the capping layer 
has a great influence on the strain in the substrate.

\section{DISCUSSION}

The experimental data on the growth of InP islands mostly involve 
InP grown on GaInP lattice-matched to a GaAs substrate.
\cite{ge95,pi96,ku95,re95,re96,ju98,pi95,ca95,an95,he96b,mo96a}  
In the cases where InP islands were used as stressors of QWs, they were 
grown on GaAs.  
In all these cases the lattice mismatch amounts to about $-3.7\,\%$. TEM 
and AFM have been used to characterize the coherent strained InP islands 
and their lateral sizes are usually found to be roughly 
\mbox{40 nm $\times$ 60 nm} and the height between 15 nm to 20 nm.

In our study we have assumed a GaP substrate with a lattice
mismatch of $-7.1\,\%$. Therefore, our data have to be rescaled 
before comparing our equilibrium shapes to the experimental observations. 
The equilibrium shape at a given volume $V$
and lattice mismatch $\alpha$ is calculated by minimizing the total energy
$E^{\rm total}(\alpha,V,I)$ with respect to the island shape $I$. As 
above, we omit the contribution from the edge energy and the 
renormalization of the surface energy due to the strain, and make use of the 
scaling behavior of the elastic energy $E^{\rm relax}$ and surface 
energy $E^{\rm surface}$,
\begin{eqnarray}
  E^{\rm relax}(\alpha,V,I) & = & 
  E^{\rm relax}_{0}(I) \: \alpha^{2} V, \\
  E^{\rm surface}(V,I) & = & E^{\rm surface}_{0}(I) \: V^{2/3},
\end{eqnarray}
where $E^{\rm relax}_{0}(I)$ and $E^{\rm surface}_{0}(I)$ are 
constants of proportionality that depend only on the shape. 
The total energy of the equilibrium island follows from
\begin{eqnarray}
  \lefteqn{ E^{\rm total}(\alpha,V)  = } \nonumber \\
   & & \ \ \ \ \alpha^{2} V \min_{I}
       \left[ E^{\rm relax}_{0}(I) + 
              E^{\rm surface}_{0}(I) \: \alpha^{-2} V^{-1/3} \right].
\end{eqnarray}
Therefore the optimum island shape is the 
same for volumes $V,V'$ and lattice mismatches $\alpha,\alpha'$,
respectively, provided that the relation
\begin{equation}
 \alpha^2 V^{1/3} = \alpha'^2 V'^{1/3}
 \label{vscal}
\end{equation}
holds. This simple relation will be altered, however, if the different 
elastic moduli for GaP and GaInP are taken into account. 

Let us select from Fig.\,\ref{fig:shapes.eps} 
the theoretical equilibrium island shape pertaining to the chemical 
potential $\mu_{\rm P}$ = $\mu_{\rm P}^{\rm bulk} - 0.1$\,eV, and volume 
$V \approx 3 \times 10^{5} \, {\rm \AA}^{3}$, because it bears the 
closest resemblance in shape to the island examined by AFM in 
Ref.\,\onlinecite{pi96}. 
We assume the volume of the latter to have the dimension given in 
Section IV, i.e., $V = 3.7 \times 10^{6} \, {\rm \AA}^{3}$. Therefore, 
for an island of volume $V \approx 3 \times 10^{5} {\rm \AA}^{3}$, 
formed with a mismatch of $-7.1\,\%$, to transform to an island of 
volume $V = 3.7 \times 10^{6} \, {\rm \AA}^{3}$, the mismatch, as given 
by Eq.\,(\ref{vscal}), ought to be $-4.6\,\%$. In view of all the 
uncertainties this appears to be reasonable. Note that the 
interdiffusion of In and Ga from the InP island and GaInP substrate, 
would decrease the assumed lattice mismatch of $-3.7\,\%$, 
hence worsening the agreement with our estimate, and would also alter 
the effective elastic moduli of the island and barrier materials. 
Although this topic has yet to be explored in these materials, there is 
clear evidence \cite{wo97} for In and Ga interdiffusion in the growth of 
${\rm In}_{\rm x}{\rm Ga}_{\rm 1-x}$As islands embedded in GaAs. A 
quantitative clarification of interdiffusion seems to be crucial for 
a better understanding of the equilibrium shape and stability of the 
strained islands. In principle, a concentration profile or phase 
separation in the island can be incorporated into the FEM calculation. 

In conclusion, we have reported an application of the 
hybrid method to study the equilibrium shape of coherent strained 
islands of InP on GaP or GaInP substrates. 
Our results are in qualitative agreement with experiments, in that all 
the observed facets are accounted for by our theory. 
We take this as an encouraging evidence that the coherent InP islands most 
often observed in experiments are close to local thermodynamic equilibrium. 
This is in contrasts to the apparent discrepancy between the calculated and 
observed island shapes for the InAs/GaAs islands.
In view of the distinct variety of observed InAs island shapes 
this discussion points towards the importance of kinetic 
effects as the missing ingredient of a more comprehensive theory for 
those islands.
The contribution of the surface energies to the total energy of an
InP island is positive, and it seems unlikely that the sign of this
contribution would change, if effects due to surface stress and strain 
were included. Thus our results favor the Ostwald ripening of 
the islands (as opposed to the creation of a thermodynamically 
stable array of islands). 
Finally, we have compared the strain fields of uncapped and capped islands. 
The capping material distinctly affects the strain fields in the 
substrate. This is of relevance for quantum dots induced in quantum wells by 
stressors.
 
\acknowledgments This work was supported in part by the Sfb 296 of the 
Deutsche Forschungsgemeinschaft.

\end{document}